\documentstyle[epsfig,epsf,psfig,multicol,aps,prl]{revtex}
\begin{document}

\title{Morphology--transitions at vicinal Cu--surfaces
based on entropic step--step interaction and
diffusion along steps}

\author{Heike Emmerich\\
{\small Max-Planck-Institut f\"ur Physik komplexer Systeme,
N\"othnitzer Str.\ 38, D 01187 Dresden, Germany}}

\maketitle
\vglue2ex
\begin{abstract}
An extension of the Burton--Cabrera--Frank model 
[Phil. Trans. R. Soc. London, Ser. A {\bf 243}, 299 
(1951)] including diffusion along steps
and entropic step--step interaction is introduced.  This extended
model is successfully applied to simulate experiments done
in the group of H.--J. Ernst at CEA Saclay, France, for vicinal
Cu surfaces.  In particular, the rise of two
qualitatively different morphologies can be explained
by the competition of growth directions
implied by the four distinct driving and restoring
forces of the model.  
Each of those forces sets
a length scale for the dominant mode of a developing
step instability. 
This explains that the wavelengthes of meandering--instabilities
observed at vicinal surfaces experimentally are usually 
larger than the one predicted by Bales and
Zangwill [Phys. Rev. B {\bf 41}, 5500 (1990)] theoretically.
\end{abstract}

\begin{multicols}{2}
During molecular beam epitaxy (MBE) appropriate
conditions for the controlled growth of 
vicinal surfaces can be fine--tuned.
This way one tries to fabricate either atomistically
flat or nanostructured surfaces.  The ability 
to exert control on structuring along the 
growth direction and fabricate substrates
with smallest--scale--built--in periodicies 
normal to the surface is well advanced.
Efforts to still enhance abilities
to determine functional properties 
of a grown substrate
now focus on the lateral structuring within 
one layer of growth.  
One direction along this line
is to make use of the inherent instabilities due to the dynamics
of the growth process itself\cite{ernst1}. 
A necessary first step is to understand the basic wavelengthes
of those inherent instabilities just as well as the 
kind of morphologies which will develop.

More then ten years ago, Bales and 
Zangwill\cite{balesz} predicted that a growing
vicinal surface should undergo a step meandering instability
when kinetic step edge barriers suppress the attachments of atoms to
decending steps\cite{ehrlich}.  According to their analysis,
a straight step is linearly unstable against perturbations
with wavelengths larger then $\lambda_c=2\pi(2  \Gamma L_\Delta)$
and a fastest growing wavelength at $\lambda_u=\sqrt{2}\lambda_c $  
($\Gamma = \frac{\Omega \gamma}{k_B T}$, $L_\Delta = \frac{D c_{eq}}{F l^2}$).
Here $F$ denotes the deposition rate,
$D$ the diffusion constant for diffusion {\sl on} the terraces 
and $c_{eq}$ the equilibrium concentration, $\gamma$ the step stiffness, $\Omega$
the atomic area and $k_B T$ the termal energy.
Even though the meandering
instability has meanwhile been observed in experiments and 
simulations \cite{emm},
%%%$jetzt korrekt oder zitate trennen auf experimentell und theoretisch
the quantitative prediction of the Bales--Zangwill analysis could
not be recovered in many 
%%%$weiss effektiv nicht wie viele$
of those experimental findings.
This point received much interest in view of the recent
experiments by Maroutian {\sl et al.}\cite{maroutian}
at CEA Saclay.  Analytical efforts to resolve 
the disagreement between experimental measurements
and theoretical prediction led to a precise study of the limit of 
desorptionless growth. For to derive a single evolution
equation in the weakly nonlinear regime this 
case has the interesting feature of displaying a singularity
in the spirit of multiscale expansion\cite{krug_long}.
As a result, meander--wavelengthes larger than the ones
predicted by Bales and Zangwill can be explained, however, without
reaching the order of experimental observations.
Other efforts center around the investigation 
of an extra diffusion current {\sl along} step edges\cite{giessen}.
Such a current triggers an asymmetry in energy barriers
for atoms attaching at kinks from different step edge directions,
the so called Kink Ehrlich--Schwoebel effect
(KESE)\cite{oplkese}. KESE can either destabilize or stabilize
steps depending on whether the slope of the meander--instability is
greater or less than one.  In \cite{maroutian} calculations including a  
stabilizing KESE--current along steps were compared to
the further thesis that islands might nucleate at step edges and
thereby constitute a still different wavelength of instability.
It appeared that the latter thesis 
%(island nucleation alongsteps) 
fits best to the experimental wavelength--values.
   
New light was shed on these investigations by experiments which revealed
that a morphology qualitatively different from the meandering one
can develop on vicinal surface as well, with growth
conditions being exactly the same as for the meandering--morphology
except a different angle of miscut\cite{maroutiantobe}.
It can be characterized by the absence of one global growth direction
and thereby resembles the so called 
{\it degenerated morphology}\cite{akamatsu}.
Thus the insufficiency of theoretical approaches to explain
the difference between the wavelength of instability predicted
by Bales and Zangwill and the one observed 
experimtentally was extended by their failure to explain this
{\it degenerated morphology}.
%\footnote{
Note that the surface of the in--phase--meandering morphology
displayed in the STM topograph in Fig. 1(b)\cite{maroutian} was wrongly identified
as Cu(1,1,17) and is in fact a Cu(0,2,24) surface\cite{tomerr}.
The true Cu(1,1,17) on the other hand is the one 
to display a {\it degenerated morphology}\cite{maroutiantobe}.
%}

In this letter I will introduce a novel model which can explain
the rise of two different morphologies as well as
their basic wavelengthes including
KESE currents plus entropic step--step repulsion
in the classical Burton--Cabrera--Frank (BCF) model\cite{bcf}.
It is the competition between the destabilizing and
the stabilizing forces combined in 
this extended BCF model (EBCF model)
which results into two different kinds of morphologies depending on the
difference in magnitude of orthogonal driving forces.
A careful analysis of this interaction also
reveals a new basic wavelength of instability
which is in the order of the experimental observations. 

To get an impression of the basic ideas underlaying the EBCF
let us start with the one sided BCF model, i.e. the model formulated
by Burton, Cabrera and Frank in 1951 for the case of complete 
supression of attachment of adatoms to decending steps.
This one--sided BCF model constitutes a {\sl moving boundary problem} 
of a diffusion--relaxation equation for the dynamics 
of the adatoms on the terraces and two boundary 
conditions for the conservation of mass 
and the conservation of energy at the steps, respectively:
\begin{eqnarray}
\partial_t c = D \nabla^2 c - \frac{1}{\tau} c + F\\
c_{eq}=c_{eq}^0\cdot(1+\kappa\Omega\gamma/k_B T)\\
v_n=D \Omega \frac{\partial c}{\partial n}. 
\end{eqnarray}
Here $c$ denotes the areal adatom density, 
%$F$ the deposition rate,
%$D$ the diffusion constant for diffusion {\sl on} the terraces and
$c_{eq}^0$ the equilibrium concentration for a straight step and
$\kappa$ the step curvature.
Equations (2) and (3) are to be evaluated at the front of each step.
%The boundary condition for the steps' back is a Von--Neumann condition.
For vicinal Cu surfaces in the temperature range of the
experiments under discussion a current involving the diffusion
coefficient $D_m$ along the kinked steps is operative\cite{giessen}. 
Following the notation of Pierre--Louis {\sl et al.}\cite{oplkese}
this diffusion along steps plus its anisotropy due to KESE
can be included in the {\sl moving boundary problem} above
extending Eq.(3) by $ - \partial_x J $,
with $ J = J_k + J_e $ ($J_k,~J_e$ as in Eq.(2)--(4) in the
paper of Pierre--Louis {\sl et al.}).
Note that unlike the authors of \cite{oplkese} I consider $\gamma$
to be the step stiffness 
instead of the line tension as the precise
measure of the step's tendency to straighten and thus reduce its
kink density.  As a consequence it is taken into account that
just as surface undergoing a faceting instability 
develops into the surface--tension--minimizing morphology, 
%which minimizes surface tension, 
%for the 
instabilities at step edges have ideal 
%preferable 
orientations which can minimize line tension as well.
This is crucial for the model since this is exactly
the point in which the Cu(1,1,17) and the Cu(0,2,24) vicinal
surface differ.  Both surfaces are made
up of (001) terraces of monoatomic hight and mean
terrace width $l=21.7${\AA}.  However, for 
the Cu(1,1,17) surface straight step edges 
before the growth of instabilities are parallel to  
the $<$100$>$ direction, whereas they are parallel to $<$110$>$
in the case of Cu(0,2,24).  As a consequence the direction
of minimal step stiffness, which can be determined to be
in $<$130$>$ via an embedded atom method (in analogy to \cite{wolf} for surfaces),
encloses an angle of appr. $63.5^0$ with the straight steps
in case of Cu(0,2,24), whereas the angle is appr. $71.5^0$ 
in the case of Cu(1,1,17).  Since growth proceeds in the
direction of minimal step stiffness, these are the orientations
for the initial formation of the instabilities.  
They trigger destabilizing $J_k$ currents
of different magnitudes along the step edge forming in this direction. 
For to evaluate $J_k$ a quantitative value for the KESE--length $L_s$
(ref. \cite{oplkese}, Eq.(3)) is required.  
The relevant kink energy barrier is the one directed in $<$$\overline{1}10$$>$,
giving 0.518eV using an effective medium theory potential\cite{reften,marko}.
The diffusion barrier for the jump of a single adatom along
the step edge in this direction is 0.399eV\cite{reften,marko}, 
leading to a KESE--length $L_s$ of 201.2.  The resulting
$j_k$ current densities for Cu(0,2,24) versus Cu(1,1,17)
differ by one order of magnitude:  for Cu(0,2,24)
$j_k$ takes a value of $7.118\cdot 10^{-7}$({\AA}$^2$s)$^{-1}$,
whereas at Cu(1,1,17) $j_k = 6.396\cdot 10^{-8}$({\AA}$^2$s)$^{-1}$.
Despite their different magnitudes in both cases KESE currents
are destabilizing, favoring unsaturating amplitude growth.
Wavelengthes of instabilities turn out to be even less
then in the BCF model.  As a consequence taking into account
KESE as only additional driving force in the BCF model is not
well suited to explain the experimental work by T. Maroutian
and coauthors.  An obvious antagonist of the destabilizing KESE
is the repulsion due to the succeeding step.
Entropic just as well as elastic interactions\cite{chaouqi} have to be taken
into account.  Since the step interaction engergy $A$ 
considering an elastic or a dipole momentum is small compared to 
the entropic repulsion in the temperature ranges under discussion\cite{williams},
it is sufficient to extend Eq.(2) by taking into account the surpression
of step wandering:
%(i.e. entorpic step repulsion):
\begin{eqnarray*}
c_{eq}=c_{eq}^0\cdot(1+ 
\frac{\kappa}{k_B T} \cdot 
%%%%(\kappa\gamma + \frac{\Omega}{k_B T} \cdot (\kappa \gamma + 
(\Omega\gamma + \frac{(\pi k_B T)^2}{6 l^2 \gamma}))~~~~~~(\tilde{2})\\
v_n=D \Omega \frac{\partial c}{\partial n}-\partial_x J~~~~~~(\tilde{3})\\
\end{eqnarray*}
The additional term 
%$\frac{\Omega(\pi k T)^2}{6 l^3 \gamma}$ is
$(\pi k_B T)^2 / 6 l^2 \gamma $ is
the step interaction parameter with $l$ equal the width of the
terraces.  Equations $(1)$, $(\tilde{2})$ and $(\tilde{3})$
constitute the extended BCF model (EBCF).
Its simulation with T=280K, F=3$\cdot10^{-3}$ ML/s, l=21.7 {\AA},
$D_m = 10^{-6}$cm$^2$/s, $c_{eq}=8.208\cdot10^{-6}${\AA}$^{-1}$  and $\gamma$= 1.034$\cdot$eV/{\AA} 
%$c_{eq}=8.2075315\cdot10^{13}ev$   and $\gamma$= 1.034273$\cdot$eV/{\AA}. 
yields the following two
morphologies for the Cu(0,2,24) surface (Fig. 1) 
and the Cu(1,1,17) (Fig. 2), respectively.

The four components regulating growth of instabilities in the EBCF are
\begin{enumerate}
\item driving of growth via the gradient of the adatom diffusion field normal to the interface
setting the length scale $L_\Delta$ for a primary wavelength of instability
\item restore via step stiffness (corresponding length scale: $\Gamma$)
\item driving of amplitude growth via KESE (length scale: $L_s$, ref.\cite{oplkese})
\item restore via entropic repulsion (length scale: $L_S=\frac{6 \gamma l^2}{k_B T \pi^2}$) .
\end{enumerate}
%%%%%%%%%%%%%%%%%%%%%%%%%%%%%%%%%%%%%%%%%%%%%%%%
%%\begin{center}
%%\begin{figure}
%%\centerline{\psfig{file=GraphCu1117.ps,width=6cm}}
%%\caption[]{\small\label{fig:arrhenius}
%%Recovery of the arrhenius law found in \cite{maroutian}
%%in simulations with the EBCF model.}
%%\end{figure}
%%\end{center}
%%%%%%%%%%%%%%%%%%%%%%%%%%%%%%%%%%%%%%%%%%%%%%%%
%%%%%%%%%%%%%%%%%%%%%%%%%%%%%%%%%%%%%%%%%%%%%%%%
%%\begin{center}
\begin{figure}
%%\makebox[0.7cm][r]{}{\sf \small 4000 GU}\vspace*{-4mm}\\
\centerline{\epsfig{figure=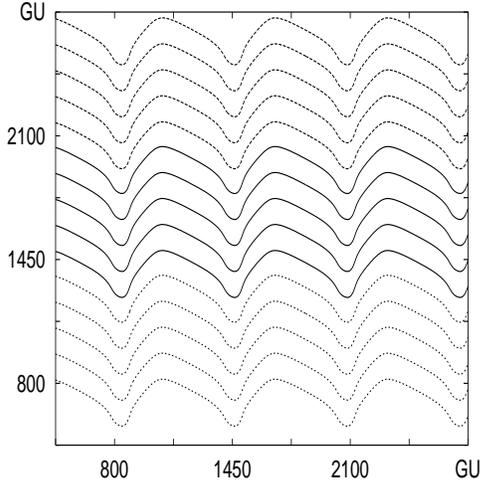,width=6.8cm,height=6.8cm}}
%%\makebox[6.5cm][r]{}{\sf \small 4000 GU}\\
\caption[]{\small\label{fig:coherent}
In--phase meandering according to numerical simulations of EBCF.
The parameters of the simulation are calibrated to Fig. 1(b)
in \cite{maroutian}. 
%The numerical grid is 2000 times 2000 grid units.
GU referes to grid units of the underlaying numerical grid.
A train of 15 steps with periodic boundary conditions in lateral direction
is displayed.  Horizontal boundary conditions are periodic
as well. Space calibration leads to roughly 10 grid units
corresponding to 1{\AA}.} 
\end{figure}
%%\end{center}
%%%%%%%%%%%%%%%%%%%%%%%%%%%%%%%%%%%%%%%%%%%%%%%%
%%\begin{center}
\begin{figure}
%%\makebox[0.7cm][r]{}{\sf \small 2000 GU}\vspace*{-4mm}\\
\centerline{\epsfig{figure=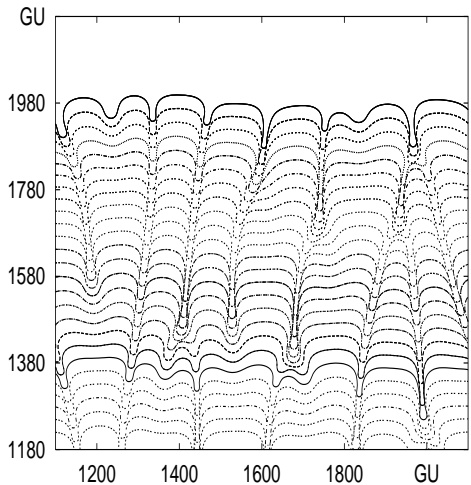,width=6.8cm,height=6.8cm}}
%%\makebox[6.5cm][r]{}{\sf \small 2000 GU}\\
\caption[]{\small\label{fig:degen}
The {\sl degenerated} morphology as obtained in simulations.
Same as in Fig.\ref{fig:coherent} except a different scale
of 1GU corresponding 1{\AA} and taking into account
the different vicinal (namely Cu(1,1,17) versus Cu(0,2,24)) 
via a different direction
of minimal step stiffness and resulting different magnitude of
KESE (refer to text).  
A train of 25 steps with first step growing 
towards infinity is displayed. Steps feel the infuence of
overlapping fields\cite{metui}.  This simulation result is in
good agreement with the {\sl degenerated} morphology recorded
for Cu(1,1,17) after deposition of about 20 ML at F=5$\cdot 10^{-3}$ ML/s
at surface temperature 285 K via STM topography\cite{maroutiantobe}.}
\end{figure}
%%\end{center}
%%%%%%%%%%%%%%%%%%%%%%%%%%%%%%%%%%%%%%%%%%%%%%%%
%%%%%%%%%%%%%%%%%%%%%%%%%%%%%%%%%%%%%%%%%%%%%%%%
%The four components regulating growth of instabilities in the EBCF are
%\begin{enumerate}
%\item driving of growth via the gradient of the adatom diffusion field normal to the interface
%setting the length scale $L_\Delta$ for a primary wavelength of instability
%\item restore via step stiffness (corresponding length scale: $\Gamma$)
%\item driving of amplitude growth via KESE (length scale: $\partial_x J_k$)
%\item restore via entropic repulsion (length scale: $L_S=\frac{6 \gamma l^2}{k_B T \pi^2}$) .
%\end{enumerate}
In the situation depicted by Fig. 2 the magnitude 
of the KESE current derivative $\partial_x J$
is neglectable small.  The remaining forces due
entropic and step stiffness
effects act in directions perpendicular to each other.
%%%%is in the order of the gradient of the adatom diffusion field
%%%%in the direction of minimal step stiffness and normal to the interface.
%%%%Due to entropic effects the forces are perpendicular and
%%%%therefore compete with each other.  
The result of this competition
of prependicular forces of equal strength is the {\sl
degenerated morphology} as observed in other contextes
with an analogous competition of driving forces\cite{akamatsu}.

In contrary in Fig. 1 the KESE current does contribute
to the evolution of surface morphology.  
The magnitude of the driving force due to this current
is of an order 10 smaller then
the magnitude of the 
adatom diffusion field gradient
in direction of minimal step stiffness.
%% is of an order 10
%% larger than the magnitude of the contributing KESE component.
The precise factor is evaluated via simulation 
and reads 11.129 for the simulation run done for Fig. 1
averaged over 500 timesteps.
Due to the dominance of $L_\Delta$ growth can proceed in the direction of
minimal step stiffness.
  
For to understand the basic wavelength of the instability  
the two new length scales $L_S$ and $L_s$, set by entropic
interaction as in Eq.($\tilde{2}$) 
and by an anisotrop diffusion current
along the step as in Eq.($\tilde{3}$) 
respectively, have to be taken into account.
Eq.(1),($\tilde{2}$) and ($\tilde{3}$) constitutes
a system with a type I bifurcation\cite{hohenberg}.
It's dispersion relation can be evaluated 
in analogy to a dilute binary alloy 
undergoing directional solidification\cite{mullins}.
For $\Gamma \ll \L_S$
%For to understand the
%basic wavelength of the instability in this case 
%KESE currents can be neglected.  However, entropic
%interaction as in Eq. $(\tilde{2})$ sets a new scale $L_S$ (item 4. above).
%The basic wavelength of instability therefore
%arises from a multiplication of the scales
%set by capillary, entropic and diffusion lengthes.
the resulting expression in leading order reads
\begin{eqnarray}
\lambda_u = \sqrt{2} \pi(\frac{L_\Delta \cdot \Gamma \cdot \L_S}{4})^{1/3}
+ 2\sqrt{2}\pi\frac{\Gamma L_s}{3 l_T} .
\end{eqnarray}
It replaces Eq.(1) in \cite{maroutian}.
Expr.(4) is sufficient in the most interesting temperature range 
$2.8$ to $3.8$ x $10^{-3} K^{-1}$. 
%%The full analytical solution for the quasistationary
%%dispersion relation, however, can only be expanded
%%in an infinite polynomial series.  Taking into account
%%second order terms results into a deviation from the
%%Arrhenius Law (Fig. 3, dotted line), which can explain
%%the change of slopes for temperature ranges
%%$T^{-1} \ge 3.6$ x $10^{-3} K^{-1}$ in Fig. 3.   
%%As a result, the wavelength now displays small
It displays deviations from an Arrhenius--type behaviour,
which increase with increasing temperature.
Nevertheless it is in good agreement
with the experimental data (Fig. 3).
Simulations of the EBCF model support this expression as
well.  
%%%%%%%%%%%%%%%%%%%%%%%%%%%%%%%%%%%%%%%%%%%%%%%%
%%\begin{center}
\begin{figure}
\centerline{\psfig{file=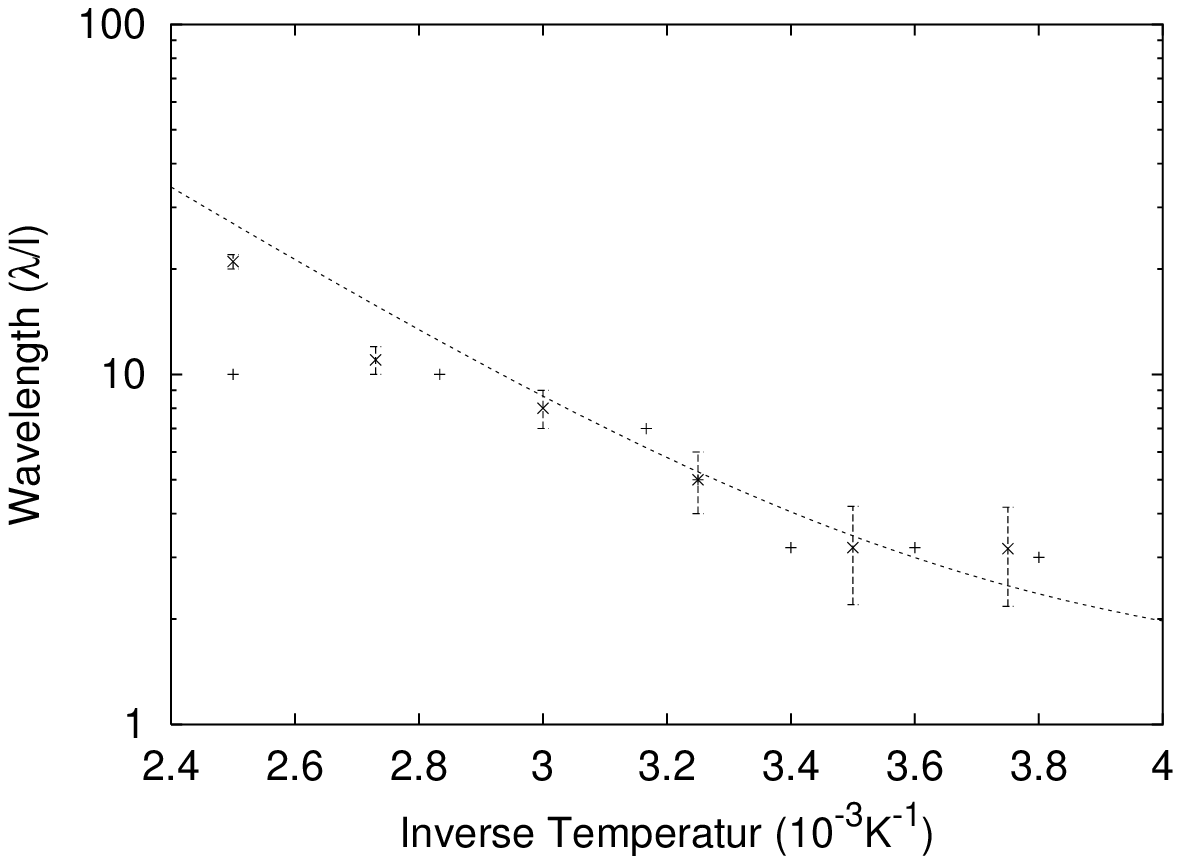,width=9.0cm}}
\caption[]{\small\label{fig:arrhenius}
The '+'--data points are taken from \cite{maroutian}
and give the wavelength measurements recorded there.
The full line is the solution of (4).  
%It does not display an exact Arrhenius--type behaviour,
%but obviously still fits the experimental data reasonable. 
Crosses with errorbars are data points obtained from simulation with the
EBCF.  There are limitations to 
a precise wavelength--measure in the simulations.  Due to the
horizontal periodic boundary conditions wavelengthes
are always a divider of the system's width.
Enlargening this system width systematically 
until the rise of a further
cell as well as restricting the width until the
extinction of one of the cells allows to find 
upper and lower bounds.  Data points are
the mean of these bounds which themselves
are taken into account via the error bars.
Thus each data point is a result of up to 15
simulations with different system widthes.
%%%%%%%%%%No fit parameters in Eq. (4).
%For all cases initial conditions were nearly straight steps.
}
\end{figure}
%%\end{center}
%%%%%%%%%%%%%%%%%%%%%%%%%%%%%%%%%%%%%%%%%%%%%%%%
%%%%%%%%%%%%%%%%%%%%%%%%%%%%%%%%%%%%%%%%%%%%%%%%
%To summarize, my extended BCF model explains two relevant
%new results, namely the rise of wavelengthes larger then
%the ones predecited by Bales and Zangwill 
%%%%%%%%%%%%%%%%%%%%%%%%%%%%%%%%%%%%%%%%%%%%%%%
%To summarize, the EBCF--model extends the original BCF--model by
%an entropic step--step interaction and an anisotrop
%diffusion stream along step edges.  This constitutes 
%a coupled set of four driving and restoring forces.
%Their competition gives rise to a new basic wavelength 
%(es sollte ja uebrigens auch in einer linearen Stabilitaetsanalyse
%rauskommen)
%of instability which is roughly seven times larger than the
%one predicted by Bales and Zangwill on the basis of the BCF--theory.
%Thereby the experimental findings
The extended BCF explains two relevant new results for growth at vicinal
surfaces, namely the experimentally observed deviation from the Bales--Zangwill
instability and the rise of a {\sl degenerated morphology}.
These results are related to experimental findings
in the group of H.--J. Ernst. 
%can be explained within the accuracy of the model.
The new basic wavelength 
predicted theoretically from the EBCF--model can be observed
as initial perturbation of the step
at Cu(0,2,24) just as well as Cu(1,1,17) surfaces.
The qualitative difference between their morphologies,
i.e. whether a surface displays an in--phase--meandering (Cu(0,2,24))
or a {\sl degenerated} structure (Cu(1,1,17)) after deposition
of a few monolayers, depends on the 
magnitude of the angle between the direction of 
destabilizing and restoring force.  
Time scales for the development from initial perturbation to
the full {\sl degenerated} morphology could not be compared
with experiments due to the lack of experimental data.
It seems an interesting question whether {\sl in situ}
transitions from one morphology to the other
can be obtained via a change in the ratio of driving forces (e.g. by lowering temperature
during the experiment). If this morphology transition
displays features which resemble a true phase transition remains to be
investigated.

The diffusion--relaxation Eq.(1) is solved on a quadratic grid.  The
interface is discretized separatly by curvilinear segments and 
respective interpolations from the interface to the grid.
Details of the scheme and its parallelization will be given in 
a further paper.  
 
I thank T. Maroutian and M. Rusanen for sending me a STM--picture of the true
Cu (1,1,17) surface--morpology as well as a respective preprint prior to publication.
Comments by H. M\"uller--Krumbhaar to focus my attention on a precise
understanding of the basic wavelength of instability arising in the EBCF--model
are gratefully acknowledged.   
Part of the computations were done on the T3E and the Origin 3800 of
the URZ at Dresden Technical University.\\
{\sl Note}--The two surfaces (Cu(1,1,17) and Cu(0,2,24))
have also been modeled in \cite{marko} on the basis of a kinetic
Monte Carlo algorithm without taking into account entropic repulsion.
The agreement with experimental findings remains restricted, nevertheless
it is a very precise study of the different effects of KESE at the different
surfaces.

\end{multicols}
\end{document}